\documentstyle[hx37_latex,psfig]{article}

\def\be{\begin{equation}}
\def\ee{\end{equation}}
\def\bea{\begin{eqnarray}}
\def\eea{\end{eqnarray}}

\def\gs{\mathrel{\raise0.35ex\hbox{$\scriptstyle >$}\kern-0.6em 
\lower0.40ex\hbox{{$\scriptstyle \sim$}}}}
\def\ls{\mathrel{\raise0.35ex\hbox{$\scriptstyle <$}\kern-0.6em 
\lower0.40ex\hbox{{$\scriptstyle \sim$}}}}

\begin{document}

\title{HST OBSERVATIONS OF DISTANT CLUSTERS: IMPLICATIONS FOR GALAXY EVOLUTION}

\author{ A.\ Dressler }

\address{Carnegie Observatories, 813 Santa Barbara Street, 
Pasadena,\\ CA 91101, USA}

\author{ I.\ Smail~$\!$\footnote{Visiting Research Associate at the
Carnegie Observatories.} }

\address{Department of Physics, University of Durham, South Road,\\
Durham DH1 3LE, England}

\maketitle

\abstracts{
The ``MORPHS'' group has completed the cataloging, parameterization,
and morphological classification of $\sim$2000 galaxies in 10 rich
clusters from $0.36 < z < 0.56$.   From a weak lensing analysis using these
data, which compares the X-ray properties ($L_X$) of the clusters with
virial temperature estimates ($T_v$) from the lensing shear strength,
we find little evidence for evolution in the $L_X$--$T_v$ relation from
that observed for local clusters. We discuss how this observation
constrains models for the X-ray evolution of clusters.  The data have
also been used to study the color dispersion of bona-fide ellipticals
in high-$z$ clusters: we find the spread to be very small, suggesting
an early formation epoch for the stellar populations of
cluster ellipticals.  This is consistent
with the evolution of the morphology-density relationship, in which we
find ellipticals to be as abundant at $z=0.5$ as in clusters today, and
already well ensconced in the dense regions.  In contrast, S0's
are less plentiful and less well-concentrated compared to the present
epoch, and spiral galaxies everywhere more abundant.  Combined with
other spectroscopic and morphological data, these observations suggest
that most of these rapidly evolving systems are not likely to become
bright ellipticals, which were more likely formed at early epochs.
Cluster S0 galaxies, on the other hand, are likely to have been produced in large numbers in the recent past.
}

\section{Introduction}

Substantial progress has been made in the study of galaxy evolution
through use of the lookback in cosmic time afforded by the observations
of distant clusters and field galaxies.  Very different histories of
star formation produce spectral and integrated color characteristics
that are essentially indistinguishable by the present epoch.
Observations of the state of galaxies at much earlier times help break
this degeneracy and offer a clearer picture of the evolutionary path
taken by galaxies of various types and masses, and in different
environments.

A group of us we call the ``MORPHS'' --- Richard Ellis, Warrick Couch, Gus
Oemler, Harvey Butcher, Ray Sharples, Bianca Poggianti, 
Amy Barger, and ourselves, has been using images from the Hubble Space
Telescope Wide Field Planetary Camera 2 (WFPC-2) and extensive ground-based
photometry and spectroscopy to study the properties of, and galaxy
populations in, rich clusters of galaxies at $z \sim 0.5$.  Here we
report on various results from our group, which should be referred to by the 
specific papers named in each section.

\section{Mass Estimates for Distant Clusters}

Rich clusters of galaxies can be identified to high redshift and can
thus be used as tracers of the evolution of structure in the universe.
Moreover, as clusters represent the extreme tail of the mass
fluctuation spectrum they provide a particularly sensitive probe of the
form of the primordial power spectrum.  One of the most widely used
techniques to identify and study the masses of distant clusters is
X-ray imaging of the hot intracluster gas bound to the cluster
potential. Published X-ray surveys indicate a reduction in the volume
density of luminous clusters at intermediate redshift.  Unfortunately,
without corroborating evidence it is difficult to determine if this
arises from a real decline in the number of {\it massive} clusters in
the past, or a change in the thermal properties of the cluster gas.
The relatively new field of gravitational lensing provides an unique
opportunity to tackle this issue by determining independent estimates of
the cluster masses, from their effects on the shapes of background
faint field galaxies.  Using the homogenous, high-quality WFPC-2
imaging obtained for distant clusters by our collaboration, we have
made a first attempt at combining X-ray and lensing observations of a
large sample of distant clusters to determine the relative importance
of cluster mass and the thermal history of the gas on the X-ray
luminosities.  The following discussion is abstracted from Smail et
al.\ (1996a).
 
For the lensing analysis we have used deep WFPC-2 imaging of 12 distant
clusters spanning the redshift range $z=0.17$--0.56.  Working from
catalogs of faint galaxies ($I_{814} = 24.0$--26.0) detected in these
fields we measure the mean shear strength --- the average, coherent
elongation of galaxy images around the cluster lens center --- within a
200$h^{-1}$ kpc aperture.  We detect the signature of
gravitational lensing in 11 of the 12 clusters; spanning nearly an
order of magnitude in lensing strength.   Moreover, the shear strength
measured on these large scales correlates well with the presence of
multiply-imaged arcs and pairs in the very central regions of the
clusters, indicating that the clusters all share similar mass profiles.  
We have examined the correlation between the cluster X-ray
luminosities and our mean gravitational shear strengths (linearly
related to the central mass and the cluster virial temperature) and
develop a model which allows us to predict the relationship expected
from the properties of local clusters.  After allowing for various
observational effects, we find that the predicted correlation is a
reasonable match to the available data (Fig.~1), indicating that there
has been little evolution in the X-ray luminosity--central mass
relationship between $z\sim 0.4$ and now.  Such limited evolution in
the X-ray luminosity--central mass relation can be reproduced by models
introducing a modest initial entropy into the gas prior to cluster
formation, possibly resulting from pre-heating by AGN or galactic
winds.  Our results demonstrate the important role weak gravitational
lensing can play in the study of the evolution of distant clusters, as
the most direct and least biased probe of their growth.

\begin{figure}
\hspace*{-3.5truein}
\psfig{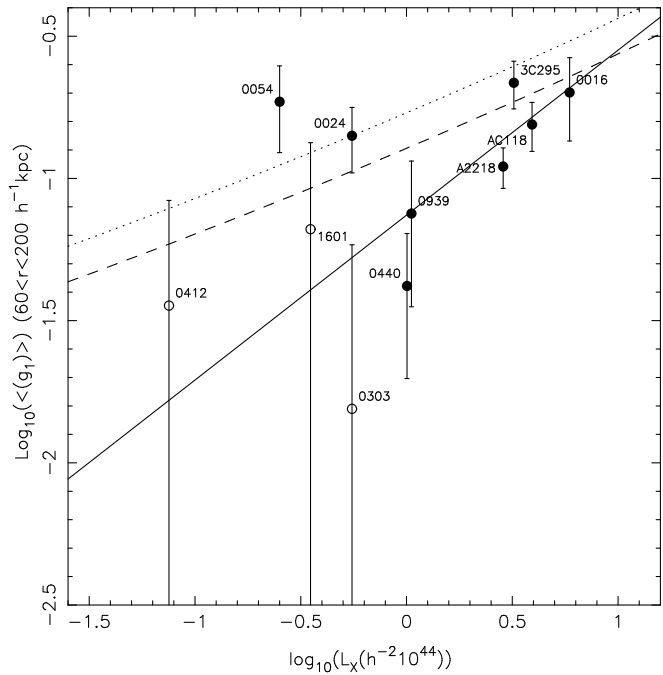}
\caption{
The correlation between the cluster X-ray luminosity and the mean shear
strength, $<\! g_1\! >$, linearly related to central cluster mass.  The
error bars are $1\sigma$ boot-strap estimates and the solid line shows
the best fit relationship for the data. The dotted line indicates the
upper limit expected, assuming 100\% measurement efficiency, in the
case of our simple model.  The dashed line represents a 75\%
efficiency.  Filled symbols denotes those clusters which have candidate
strongly-lensed features.\hfill~
}
\end{figure}

\section{The Ages of Elliptical Galaxies in Distant Rich Clusters}

Elliptical galaxies are conventionally regarded as old galactic systems
whose star formation history can be approximated as a single burst that
occurred 12--16 Gyr ago. However, in recent years, this simple picture
has been challenged from various viewpoints. Numerous cases have been
found of ellipticals with intermediate-age stellar populations and
dynamical arguments suggest that many peculiarities seen in ellipticals
(shells and dust-lanes) are best explained via recent formation from
the merger of gas-rich systems.  Nevertheless, the small scatter
observed for the $(U-V)$ colors of spheroidal galaxies in nearby
clusters of galaxies still provides a basic constraint on the history
of star formation in dense environments (Bower, Lucey \& Ellis 1992).
Using these local data and assuming the spheroidals formed
stochastically, it is possible to limit their formation epoch to $z\geq
2$.  However, introducing some degree of synchronicity serious weakens
this limit, allowing the spheriodal population to form at more recent
epochs.  

In Ellis et al.\ (1996) we address this ambiguity using high precision
rest-frame $(U-V)$ photometry of a large sample of
morphologically-selected spheroidal galaxies in three $z\sim0.54$
clusters which have been observed as part of our HST program.  We use
our $F555W$ and $F814W$ imaging to determine accurate rest-frame
$(U-V)$ colors for spheroidal galaxies in the three clusters:
Cl0016+16 ($z=0.55$), Cl0054$-$27 ($z=0.56$) and Cl0412$-$65
($=0.51$).  Using these new data we repeat the color-scatter analysis
conducted locally at a significant look-back time.  Matching our
aperture sizes, luminosity range and color system to those used locally
we find a small scatter ($\leq 0.07$ mag rms, not much greater than that
observed at $z\sim 0$) for galaxies classed as Es and E/S0s, both internally
within each of the three clusters and externally from cluster to
cluster. We do not find any trend for the scatter to increase with
decreasing galaxy luminosity beyond that due to observational error. Our
result thus provides a new constraint on the star formation history of
cluster spheroidals prior to $z\simeq0.5$. Although we cannot rule out
the continued production of {\it some} ellipticals, our results do
indicate that the bulk of the stars seen in luminous elliptical cluster
galaxies were formed by $z\simeq3$.

\section{The Morphology-Density Relation at High Redshift}

A principal goal of our group has been to study the evolution of
morphological types in the rich cluster environment.  To this end we
have morphologically classified 1857 objects brighter than $R_{702} < 23.0$
or $I_{814} < 23.5$, in the 11 fields, as described in Smail et al.\ (1996b)
While the addition of spectroscopically-derived parameters, such as
cluster membership or stellar population, is important for
understanding the evolutionary state of these populations,
photometric/morphological information alone allows a simple and
important comparison with the properties of present day clusters.  This
comparison offers clues as to how clusters of galaxies came to hold
their atypical complements of galaxy types.

With a resolution approaching 0.1$''$, our WFPC-2 images show detail at
the level of 500 pc in the clusters, equivalent to observing galaxies
in the Coma cluster with 1$''$ seeing.  While cruder than than the
resolution usually available for morphological classification of nearby
galaxies, it is sufficient for the identification of basic
morphological information, and, in particular, is comparable to that
presented by Dressler (1980) in the study of galaxy morphology in 55
low-redshift clusters.  Our morphological samples in the distant
clusters have also been selected in as similar a fashion as possible to
the local data, for example, over the same area and to comparable
absolute magnitude limits.  

Dressler found a strong relation between the fractions of E, S0, and
spiral galaxies with the local projected density where they were found,
in the well known sense that ellipticals became more prevalent, and
spirals less so, in regions of higher surface density.  He concluded
that, to first order at least, the morphology-density relation is 
universal, that is, representative of every cluster in the sample, 
regardless of its global properties. 

Dressler's original data have been reanalyzed and are presented in
Dressler et al.\ (1996) along with the morphology-density relation for
the HST sample described here.  That paper also reviews some of the
challenges to the morphology-density relation, for example, the
contention by Whitmore, Gilmore, \& Jones (1993) that the principal
determinant of galaxy type within rich clusters is the radial distance
from the cluster center.  For our purposes here we simply analyze the
morphology-density relation in our $z \sim 0.5$ sample.

In Fig.~2 we show the morphology-density relation for
the entire $z \sim 0.5$ sample.  The density range encompassed by the
more distant sample is shifted by half a dex to higher density,
probably reflecting the fact that these clusters are systematically
richer than the typical clusters of Dressler's local sample.

Before addressing the question of gradients in Fig.~2, we take note of
differences between this and the nearby cluster sample.  As is
now well known, spirals are greatly overabundant at these high
densities compared to present-epoch clusters, but, perhaps
surprisingly, the difference seems to made up entirely by a paucity of
S0 galaxies rather than an underabundance of {\it both} S0 and E
galaxies.  In fact, E galaxies appear to be in even greater abundance!
At comparable densities, spirals are a factor of 2 overabundant, S0's
are a factor of 2--3 underabundant, and ellipticals are a factor of 1.5
overabundant in the $z \sim 0.5$ sample compared to Dressler's sample
of nearby clusters.  The paucity of S0 galaxies is particularly
noteworthy. As explained in Smail et al.\ (1996b), we have compared the
distribution in flattenings of S0 galaxies in our sample with that for
the Coma cluster, to see if we have systematically misclassified S0
galaxies as ellipticals, particularly for the face-on cases.  The good
agreement of these distributions indicates that this is probably not the case,
but at any rate it is hard to see how we could be missing more than
$\sim25$\%, which is of little consequence to the gross deficiencies in
S0's found here.

\begin{figure}
\centerline{\hbox{\psfig{figure=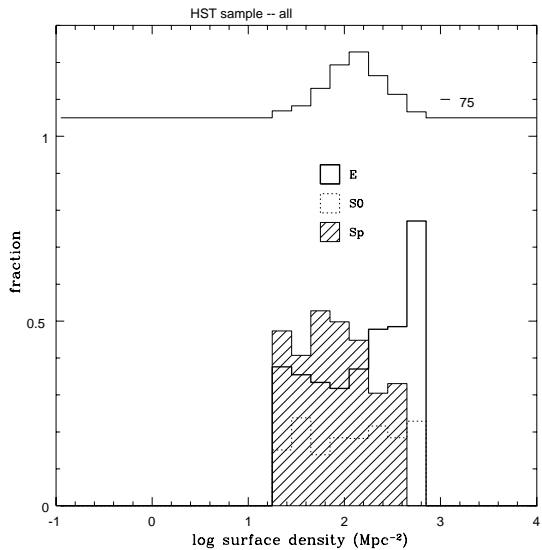,width=3.0in}}}
\caption{The morphology-density relation for 10 clusters at 
redshifts $0.36 < z < 0.57$. The plot shows the relative proportions
of the different morphological classes (E, S0, Sp) as a function of projected
density.  The upper histogram shows the total number of galaxies in
each bin. \hfill~ }
\end{figure}

We now ask whether any trend of morphology with density is apparent for
the distant sample.  From Fig.~2 it appears that a modest relation is
present, but it is only for the bins of highest surface density ---
over the last factor of 5 in surface density.  Over this range the
spiral fraction plummets and the elliptical fraction rises sharply, but
for the lower density zones, over which there is a very noticeable
gradient in the nearby clusters, the relationships are basically flat.
However, when the sample is divided by concentration and the degree of
regularity, which to a large extent go together, a very different
picture emerges.  Fig.~3 shows that, for the 4 highest concentration,
regular clusters of the $z \sim 0.5$, the morphology-density relation
is steep and well defined over the entire density range.  In contrast,
however, there are no correlations at all for the 4 lowest
concentration, irregular clusters.  This is a strikingly different
result from the situation for present epoch clusters, for which
Dressler found a strong morphology-density relationship for both
irregular and regular clusters.  A more detailed discussion of this
difference can be found in Dressler et al.\ (1996).

\begin{figure}
\centerline{\hbox{\psfig{figure=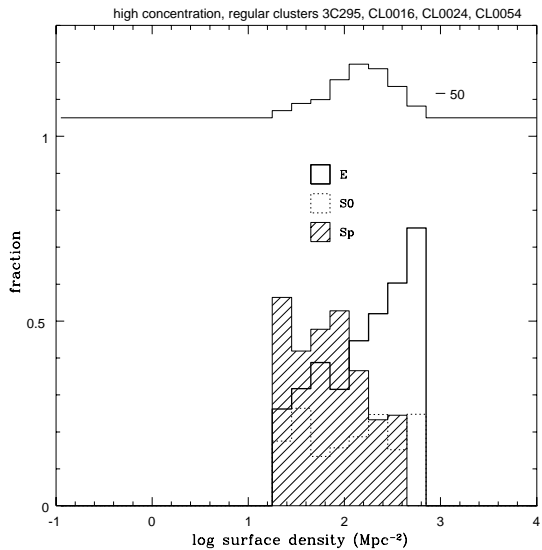,width=3.0in}}}
\caption{The morphology-density relation for 4 high-concentration, 
regular clusters at intermediate redshift, 3C295, Cl0016+16, Cl0024+16, 
and Cl0054$-$27.\hfill~ }
\end{figure}

Perhaps our most important result of this analysis, however, is simply
that irrespective of whether the clusters appear dynamically ``mature''
or not, the incidence of elliptical galaxies is already very high, and
independent of whether they are collected into dense, central regions
or not.  We suggest, based on this result, that elliptical galaxies
predate, and are basically independent of, the virialization of a rich
cluster.  This is, of course, consistent with the Ellis et al.\ (1996)
result described above of early formation of the stars in elliptical
galaxies.  Furthermore, we find that that the fractional representation
of highly asymmetric or disturbed morphologies ($D > 1$, see Smail et
al.\ 1996b) with local surface density mirrors the S0 or spiral trend
rather than the trend for ellipticals.  Together, these three
observations suggest that, for the environments of rich clusters at
least, the stars seen in elliptical galaxies are not the result of
mergers of starforming, gas-rich systems after a redshift $z = 3$.
This does not preclude the possibility of dissipationless mergers at $z
= 1$, say, when these clusters might have been in the process of
amalgamating small groups with lower velocity dispersion, but both the
distribution and numbers of ellipticals we have found here, and their
photometric and spectral properties, suggest that the stellar
populations of ellipticals in these regions are not produced by late
mergers, or in any process that depended on the dynamical evolution of
a rich cluster.  Instead, gaseous mergers or coherent collapse at high
redshift, or growth of spheroids through dissipationless mergers until
later epochs, seems to be the history indicated for ellipticals.  It is
remarkable, we think, that the environment of proto-clusters of this
richness was able to produce such a large population of ellipticals
before the identities of the clusters themselves was well established.

The situation for the S0 galaxies seems to be just the opposite.
Though the ones we find are, like the ellipticals, red and with little
scatter in color, their numbers are so deficient as to suggest that
many need to be added since $z = 1$, in order to reach the populations
of present-epoch clusters.  The source of these S0's seems clear: the
overabundance of spirals provides a resevoir of galaxies which may be
stripped by ram pressure, tidally harassed (Moore et al.\ 1995) merged,
or subject to strong 2-body gravitational interactions, with the result
of producing today's dormant disk galaxies in clusters.  Our $z \sim
0.5$ cluster sample includes a significant number of disturbed,
distorted morphologies, often with spectroscopic evidence of strong
episodes of star formation.  These may be the result of mergers, strong
interactions, accretions, harassment, or stripping --- we are still
unable to tell which of these processes are responsible.  But, we do
know from our morphological classifications that most of these are {\it
disk} systems --- they do not seem destined to settle into ellipticals
galaxies when their jostling and bursts of star formation have ceased.
Though the exact mechanism(s) may be yet unspecified, it seems that at
least half of the S0 galaxies in today's clusters have been made by
such processes since $z = 0.5$

\section{Conclusions}

Our HST images of distant clusters exhibit robust shear fields due to
gravitational lensing. These allow us to estimate the cluster mass 
(or virial temperature) and compare these to the cluster X-ray luminosity.
The relation between $L_X$ and $T_v$ in these distant clusters is
similar to that observed locally and we thus claim that there is
no strong evolution in the $L_X$--$T_v$ relationship out to $z\sim 0.4$.

We find a remarkably small scatter of restframe $(U-V)$ color for
galaxies that we have classified as ellipticals.  This suggests an
early epoch of formation for the stars in these galaxies, and perhaps
of the galaxies themselves.  Furthermore, the large number of
elliptical galaxies in these clusters, $\sim$40\%, suggests that the
formation of ellipticals predates cluster virialization.  If mergers
are responsible for making the ellipticals that now inhabit these rich
clusters, they must have been dissipationless, in the ``group phase''
at $z \sim 1$, or much earlier, $z > 3$, if significant dissipation and
star formation were involved.  In contrast, the relative paucity of
S0's in the intermediate redshift clusters suggests that many of them
have indeed been added since $z \sim 0.5$, by mechanisms that acted on the
excessive numbers, compared to today's clusters, of spirals and
irregulars.

\end{document}